# Low-complexity Point Cloud Filtering for LiDAR by PCA-based Dimension Reduction

Yao Duan, Chuanchuan Yang, *Senior Member, IEEE,* Hao Chen, Weizhen Yan and Hongbin Li

This work was supported by National Key R&D Program of China under Grant No. 2019YFB1802904. (Corresponding author: Chuanchuan Yang.)

Y. Duan, C. Yang and H. Li are with the State Key Laboratory of Advanced Optical Communication Systems and Networks, Peking University, Beijing 100871, China (e-mail: duan.yao@pku.edu.cn, yangchuanchuan@pku.edu.cn, lihb@pku.edu.cn ).

H. Chen and W. Yan are with the SenseFuture Technologies Co., ltd, Beijing 100025, China (e-mail: chenhao@sensefuture.cn, yanweizhen@sensefuture.cn ).

*Abstract*—Signals emitted by LiDAR sensors would often be negatively influenced during transmission by rain, fog, dust, atmospheric particles, scattering of light and other influencing factors, causing noises in point cloud images. To address this problem, this paper develops a new noise reduction method to filter LiDAR point clouds, i.e. an adaptive clustering method based on principal component analysis (PCA). Different from the traditional filtering methods that directly process three-dimension (3D) point cloud data, the proposed method uses dimension reduction to generate two-dimension (2D) data by extracting the first principal component and the second principal component of the original data with little information attrition. In the 2D space spanned by two principal components, the generated 2D data are clustered for noise reduction before being restored into 3D. Through dimension reduction and the clustering of the generated 2D data, this method derives low computational complexity, effectively removing noises while retaining details of environmental features. Compared with traditional filtering algorithms, the proposed method has higher precision and recall. Experimental results show a F-score as high as 0.92 with complexity reduced by 50% compared with traditional density-based clustering method.

## I. Introduction

LiDAR is a high-precision sensor used to measure the location and shape of objects and form high-quality 3D point cloud images [1, 2] which has been widely applied to autonomous driving [3, 4], 3D reconstruction [5-7], industrial measurement [8, 9] and many other areas [10].

Due to adverse effects of atmospheric particles, multipath returns caused by diffuse reflection [11], as well as adverse weather conditions such as rain and snow [12], the point cloud image obtained by LiDAR sensors suffers a lot of noises, which fall into three main categories: isolated outliers, clustered noises, and noise points near signals. In order to obtain high-quality point cloud images, these noises have to be removed.

To this end, many point cloud filtering algorithms have been developed. Rusu et al. proposed statistical outlier removal (SOR) filter which differentiates signals from noises by calculating the average distance between each point and its nearest K points [11]. Jiang and his colleagues proposed the spatial frequency (SF) outlier filter which constructs a ball with each point as the center and removes the noise points based on the number of points within the ball [13]. The above statistical outlier removal filters cannot identify outlier groupings, and the radius outlier removal filters may falsely remove a lot of useful environmental features [12]. The cell histogram filter proposed by Carrilho et al. [11] and a dynamic radius outlier removal filter proposed by Charron et al. [12] were attempts to address these problems. But the performance of radius outlier removal filter was only tested in a moderately snowy setting. Wang also proposed an adaptive ellipsoid searching filter based on the previous radius filter [14]. The center of the ellipsoid is the target point to be detected. The noises are identified based on the number of nearby points in the ellipsoid. However, this algorithm is too complex to have ideal real-time performance.

Ullrich et al. reported that noises induced by the presence of particles in the air, ground obstacles and systematic errors of LiDAR can be removed by analyzing the density of cloud points [15]. Later, Hui et al. proposed a denoising algorithm based on empirical mode decomposition, which can automatically detect outliers by identifying the components dominated by noises [16]. Yet, this method is only applicable to airborne lidar and Ulrich et al. argued that the noise in airborne LiDAR point cloud data needs to be processed by spatial averaging, but the denoising effect still is not ideal [17].

Traditional filtering algorithms are often limited to isolated outliers. In the areas close to the sensor, noise points tend to gather together and traditional distance-based filtering algorithms are not applicable in such scenarios. Meanwhile, in the areas far away from the sensor, the signal points are often so sparse that important data on environmental features often fail to be retained by traditional methods, resulting in low precision and recall of the overall point cloud. Therefore, it is necessary to introduce an unsupervised machine learning method to remove noise points that cannot be filtered by traditional methods, while preserving features.

Clustering is an unsupervised machine learning technique. There are two common types of clustering-based filtering algorithms: distance-based clustering algorithm and density-based spatial clustering algorithm. Since the former is vulnerable to noises, the density-based spatial clustering algorithm is often used to denoise 3D point cloud images [18, 19].

Ni et al. proposed a point cloud filtering algorithm combining clustering and iterative graph cuts to classify and process the point cloud data captured by airborne lidar [20]. Kim et al. proposed a graph-based spatial clustering algorithm for better segmentation of point clouds while reducing the background noise of each cluster [21]. And density-based spatial clustering of applications with noise (DBSCAN) is a widely used clustering technique, which is very effective at noise-reduction [22]. The problem is that all the three methods above are often accompanied by too high complexity to fulfill real-time tasks [20-22].

Different from all the filtering methods above which directly process 3D point cloud data, this paper proposes a new noise-reduction method which can generate low algorithm complexity through dimension reduction based on principal components analysis (PCA), while retaining environmental features. We refer to the proposed method as the PCA-based adaptive clustering filtering method (PCAAC). The innovation points of this method are as follows:

1) The proposed method employs PCA technology for dimension reduction (3D→2D) and generates 2D data by extracting the first principal component and the second principal component of the original data with little information attrition. Since main signal processing is executed on 2D data, this method can significantly reduce the overall complexity while effectively removing noises and preserving environmental features.
2) We propose an adaptive clustering method for 2D point cloud before they are restored into 3D. Through adaptive parameter setting on 2D space spanned by two principal components, the precision and recall of point cloud images are significantly improved.
3) To address the near-far effect caused by distance differences from the viewpoint to scanned points, the proposed method adopts a different region segmentation approach, which can help yield better noise-reduction effects.

Compared with the density-based clustering algorithm, the method proposed in this paper has higher accuracy and recall, with the F-score increased to 0.92 and the complexity reduced by as much as 50%. Compared with the traditional filtering algorithms such as the radius outlier removal filter, the proposed method has higher performance, with the F-score up by 50% and the complexity reduced by three times.

The rest of this paper is organized as follows: Section 2 lays out the principle of the proposed method. Section 3 analyzes the complexity of each filtering algorithm. Section 4 describes the filtering performance. Finally, conclusions are drawn.

## II. Low-complexity Point Cloud Filtering by PCA based Dimension Reduction

### A. PCA based Dimension Reduction for LiDAR

PCA is a technology that can extract main feature components of data [23, 24]. The eigenvector corresponding to the smallest eigenvalue is often noises related and discarding this eigenvector can help with noise-reduction. Assuming that the number of 3D data points is $m$, these data points are first arranged into matrix $\xi$ with 3 rows and $m$ columns. Each row of matrix $\xi$ is zero-centered by subtracting the means, which can ensure that the average value of each row is zero. Then the covariance matrix $C$ is obtained by

$$C = \frac{1}{m}\xi\xi^T \quad (1)$$

After that, three eigenvectors are arranged into matrix $E = (e_1, e_2, e_3)$ and the following equation must be satisfied:

$$E^T C E = \Lambda \quad (2)$$

where $\Lambda$ is a diagonal matrix. Therefore, three eigenvectors and their corresponding eigenvalues are identified. The eigenvectors are arranged into a matrix in rows with decreasing eigenvalues, and the first two rows of eigenvectors are taken to generate matrix $P$. The third row contains little information and is thus removed. The original 3D data can be converted into 2D as:

$$Y_{2\times m} = P_{2\times 3} \times \xi_{3\times m} \quad (3)$$

There are altogether $m$ new 2D data points, each composed of only the first principal component and the second principal component. Converting the point cloud data from 3D to 2D can reduce computational complexity of subsequent signal processing algorithms.

Based on this theory, this paper proposes an adaptive clustering filtering method. Different from the traditional filtering methods that directly process 3D point cloud data, the proposed method converts the original 3D data into 2D through PCA and then clusters the generated 2D data for noise reduction before restoring the data into 3D.

### B. Adaptive clustering filtering method based on PCA

The proposed PCA-based clustering filtering method is outlined in Fig. 1. The original data are first divided into region segments in response to the decreasing density of point cloud further away from the sensor. PCA is then carried out on each region for dimension reduction following the method in Section A. The first principal component and the second principal component, whose accumulative variance contribution rate exceeds 95%, are extracted and designated as X-axis and Y-axis coordinates. Next, the 2D clustering algorithm is used to cluster the planar point clouds. The numbers of point clouds in each cluster are calculated and compared with the threshold to identify and remove noise points. Afterwards, the filtered 2D point cloud data are restored into 3D point cloud data. Finally, the restored 3D data for all segments will be stitched together to form a complete point cloud image. Each step will be elaborated in the following sections.

#### 1) Region Segmentation Approach

The distance from each area to the sensor is different, and the number of buildings and street scenes are also varied, requiring different filtering parameters. By segmentation, data points in different areas will be filtered with different parameter settings based on their respective environmental features.

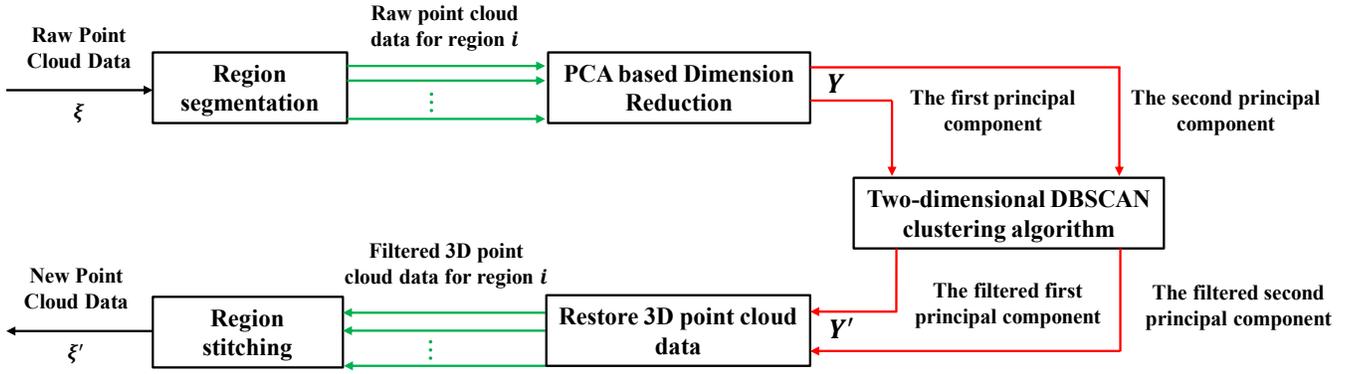
Fig. 1. The block diagram of the proposed PCAAC method.

Fig. 2(a) shows the traditional segmentation method with the LiDAR sensor at the center. The whole space is divided into multiple cube regions with a same volume. However, the distances from the points within the same cube to the sensor are different, and the noise distribution is also different. For example, $d1 \neq d2 \neq d3$ and $d4 \neq d5$, which goes against the principle of segmentation. Therefore, we propose an improved segmentation approach, as shown in Fig. 2(b).

Given that real-life environment is mainly composed of buildings and streets with limited heights, Z-values in Cartesian coordinate system usually range between 0 and 30 meters. While the height of an object is within a certain range, the ground (X-Y plane) is theoretically indefinite. Therefore, we propose a cylinder-structure segmentation approach. As shown in Fig. 2(b), with the LiDAR sensor at the center, the whole space is segmented into many cylinder blocks, and the space sections between each cylinder are the segment regions of point cloud.

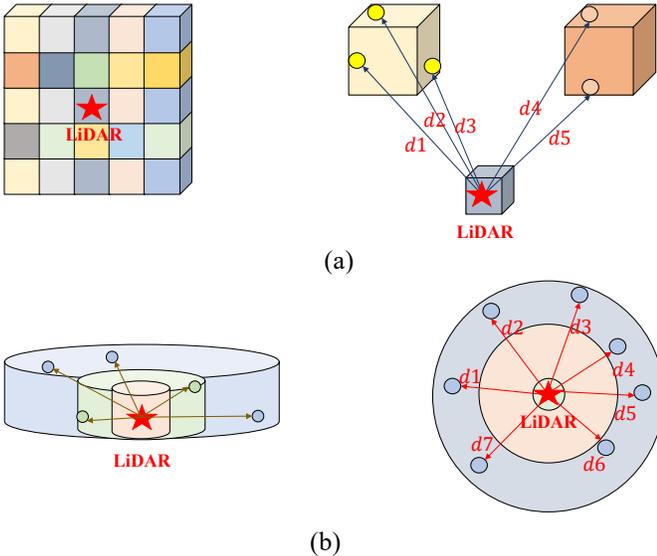

(b)

Fig. 2. Traditional (a) and improved (b) region segmentation methods.

In this cylinder model, to have a same volume for each region, the bottom area between every two neighboring cylinders should satisfy

$$\pi r_1^2 h = \pi r_2^2 h - \pi r_1^2 h = \cdots = \pi r_t^2 h - \pi r_{t-1}^2 h. \quad (4)$$

where $r_i$ ($i=1, 2\ldots, t$) is the radius of the $i$th cylinder and $h$ is the height of cylinder. This formula can be simplified into $r_i = \sqrt{i} r_1$, meaning when the bottom radius of cylinder $i$ is $\sqrt{i}$ times of that of cylinder 1, the volume of each segment is equal. $r_1$ is determined by the point furthest away from the sensor as

$$r_1 = \frac{r_{max}}{\sqrt{t}} \quad (5)$$

After segmentation, PCA is performed on the raw 3D data to extract main components and convert 3D Data into 2D Data.

*2) 2D-DBSCAN Clustering*

Based on the theory of dimension reduction in Section A, a 2D-DBSCAN clustering algorithm is proposed to filter 2D point cloud on a plane. By calculating the number of point clouds in each cluster and comparing it with the threshold, we can identify noises. A cluster with fewer point clouds than the threshold are identified as a signal cluster; otherwise, it is identified as a noise cluster.

The DBSCAN algorithm can find all the dense regions of the sample points and treat these dense regions as clusters. Therefore, signal points and noise points can be distinguished as shown in Fig. 3. It has two algorithm parameters $\varepsilon$ and $minpts$, where $\varepsilon$ is the specified neighborhood radius and $minpts$ is the minimum number of points within the $\varepsilon$ neighborhood of a core point.

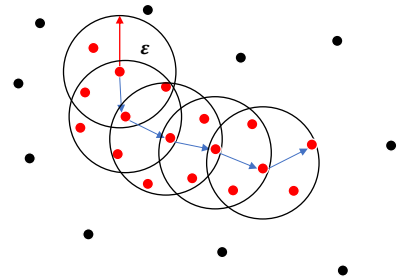

Fig. 3. Clustering results of DBSCAN algorithm. Points within a cluster (signal point) are marked red; points outside the clusters (noise point) are marked black.

The original 3D point cloud is represented by an unordered set of points

$$\xi = \{(x_i, y_i, z_i)^T, i = 1, 2 \ldots m\} \quad (6)$$

where $(x_i, y_i, z_i)^T \in R^3$ is a coordinate vector and $m$ is the total number of points.

According to formula (3), the original 3D data can be converted into 2D data after PCA dimension reduction. Each 2D point cloud is composed of the first principal component and the second principal component, represented by $f$ and $s$ respectively.

Therefore, the point cloud input into DBSCAN cluster can be expressed as

$$Y = \{(f_i, s_i)^T, i = 1, 2 \ldots m\} \quad (7)$$

where $(f_i, s_i)^T \in R^2$ is a coordinate vector in the 2D space spanned by two principal components.

---
**Algorithm 1** 2D-DBSCAN Algorithm
---
**Input:** $Y$, $\varepsilon$, $minpts$, $d$, $\psi$
**Data:** $label$
**Function:** $RANGE\_QUERY(\cdot)$
1: **for** each point $p_i = (f_i, s_i)^T$ in $Y$ **do**
2:     **if** $label_{p_i} = undefined$ **then**
3:         Neighbors $N \leftarrow RANGE\_QUERY(Y, d, p_i, \varepsilon)$
4:         **if** $|N| < minpts$ **then**
5:             $label_{p_i} \leftarrow Noise$
7:         **else**
8:             $S \leftarrow$ new cluster
9:             Seed set $S \leftarrow N$
10:         **end if**
11:     **end if**
12:     **for** each point $p_{i,j} = (f_{i,j}, s_{i,j})^T$ in $S$ **do**
13:         **if** $label_{p_{i,j}} = undefined$ **then**
14:             Neighbors $N' \leftarrow RANGE\_QUERY(Y, d, p_{i,j}, \varepsilon)$
15:             **if** $|N'| \geq minpts$ **then**
16:                 $S \leftarrow S \cup N'$
17:             **end if**
18:         **end if**
19:     **end for**
20: **end for**
21: **Output:** $\{S_1, S_2 \ldots S_k\}$
22: **for** each cluster $S_i$ in data set $\{S_1, S_2 \ldots S_k\}$
23:     **if** $|S_i| < \psi$ **then**
24:         **for** each point $p_c = (f_c, s_c)^T$ in $S_i$ **do**
25:             $label_{p_c} \leftarrow Noise$
27:         **end for**
28:     **end if**
29: **end for**

---

The 2D-DBSCAN algorithm is displayed in *Algorithm 1*.

$Y$ refers to the data set composed of the first principal component $f$ and the second principal component $s$. Specifically, in the 2D plane, the distance $d$ can be calculated as:

$$d = \sqrt{(f_1 - f_2)^2 + (s_1 - s_2)^2} \quad (8)$$

After being visited, for every unvisited point $p_i$ in data set $Y$, identify all points within its $\varepsilon$ neighborhood with $RANGE\_QUERY$ function and pool into a subset $N$. The number of points within subset $N$ is referred to as $|N|$. If $|N| < minpts$, point $p_i$ is labeled as an outlier; if $|N| \geq minpts$, point $p_i$ is a core point and a new cluster $S$ will be established where subset $N$ is added. For an unvisited point $p_{i,j}$ in $S$, if the number of points within $N'$ equals to or exceeds $minpts$, $N'$ will be added to $S$.

The above process is repeated until there is no unvisited point, resulting in $k$ clusters: $\{S_1, S_2 \ldots S_k\}$.

The number of data points within cluster $S_i$ ($i = 1, 2 \ldots k$) is referred to as $|S_i|$. Compare $|S_i|$ with the threshold $\psi$. If $|S_i| > \psi$, cluster $S_i$ is a signal cluster. Otherwise, it is identified as a noise cluster and removed. The threshold $\psi$ of each area is different, depending on the distance from the lidar sensor. The farther away from the sensor, the larger the signal block obtained by clustering, and thus the larger the threshold should be correspondingly.

After removing the noises, we can get the filtered point cloud data $Y'$, which can be expressed as:

$$Y' = \{(f_l, s_l)^T, l = 1, 2 \ldots n\} \quad (9)$$

where $(f_l, s_l)^T \in R^2$ is the filtered coordinate vector in the 2D space and $n$ is the number of points remaining after filtering.

After being denoised, the 2D data is restored into 3D. The denoised 3D data can be obtained through:

$$\xi' = P^T Y' + \bar{\xi} \quad (10)$$

where $Y'$ is the filtered point cloud data set in 2D space; $\bar{\xi}$ is the means subtracted from the original matrix $\xi$.

### C. Key Parameters

There are three key parameters in the proposed PCAAC method, i.e. $\varepsilon$, $minpts$ and $\psi$. For each region segmentation, the three parameters are adaptively modified to obtain the best clustering effect and high-precision point cloud data.

Considering that the density of cloud points declines further away from the sensor, the $\varepsilon$ for each region segment is modified adaptively. As shown in Fig. 4, the closer the point cloud is to the LiDAR sensor, the higher the density, and the smaller the $\varepsilon$ value should be.

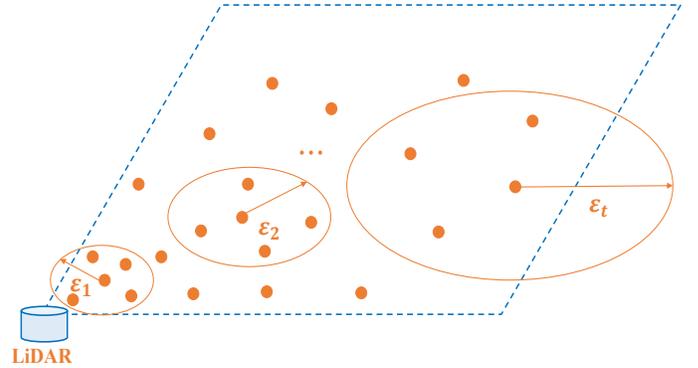

Fig. 4. Adaptive $\varepsilon$ parameters for each region segmentation.

$\varepsilon$ for a certain region can be defined as:

$$\varepsilon_i = k r_i. \quad (11)$$

where $\varepsilon_i$ ($i = 1, 2 \ldots, t$) refers to the neighborhood radius for region $i$ and $r_i$ refers to the distance between the point cloud in region $i$ and the lidar sensor. Since $r_i = \sqrt{i} r_1$, there is:

$$\varepsilon_i = \sqrt{i} \varepsilon_1 \quad (12)$$

where $\varepsilon_1$ is equal to 1.

The default value of $minpts$ is set as 10. Based on $\varepsilon$ and $minpts$, silhouette value is calculated. The value of $minpts$ for each region is adjusted until the largest silhouette value is obtained.

$\psi$ is used to distinguish noise clusters from signal clusters. Given that the number of point clouds for a general object is more than 100, the default value of $\psi$ is set at 100. The threshold range is between 50 and 200.

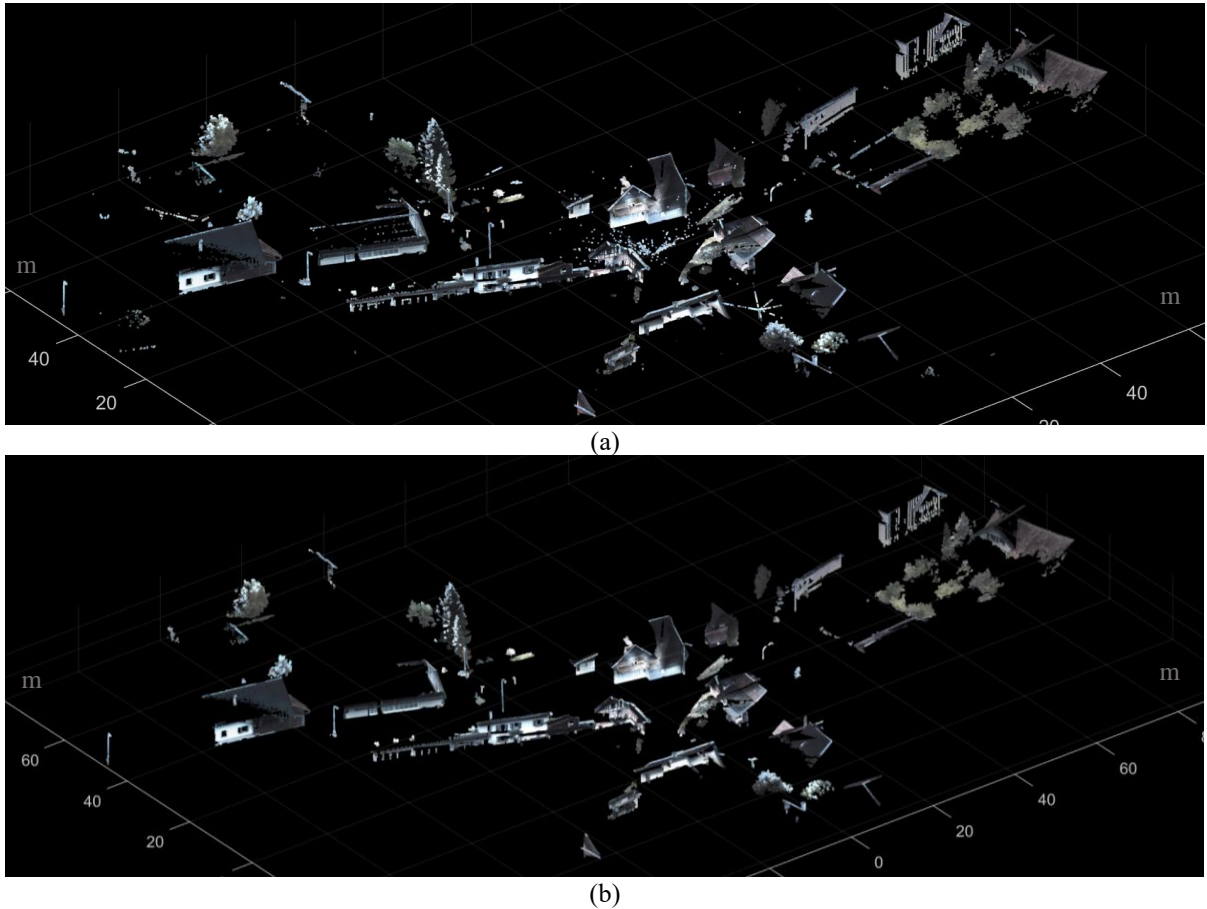

Fig. 5. (a) Original point cloud image, (b) Filtered point cloud image by the proposed PCAAC filter.

## III. COMPLEXITY ANALYSIS

Table I compares complexities of different point cloud filtering methods. Compared with the statistical outlier removal filter, the complexity of the proposed PCAAC method is reduced by an order of magnitude. Compared with the radius outlier removal filter, the complexity of PCAAC method is reduced by three times. Compared with the existing density-based clustering method, the complexity of PCAAC method is reduced by about 50%. The complexities listed in *Table I* are calculated when the corresponding filtering methods yield the best performance and their performances are analyzed in the following Section.

TABLE I
THE COMPLEXITY ANALYSIS OF DIFFERENT METHODS

| Filter | Complexity (+) | Complexity (×) |
|---|---|---|
| *Statistical outlier removal filter* | $O(m^2 \log m)$ | $O(m^2 \log m)$ |
| *Radius outlier removal filter* | $O(9m^2)$ | $O(3m^2)$ |
| *Density-based clustering method* | $O(13m^2)$ | $O(1.5m^2)$ |
| *PCAAC method* | $O(10m^2)$ | $O(m^2)$ |

## IV. PERFORMANCE ANALYSIS

The original point cloud data used in performance analysis came from global public database *Open Topography*. The point cloud processing software is MATLAB provided by MathWorks.

Fig. 5(a) shows the original point cloud data, including buildings, grasslands, trees, street lamps and other environmental features. There are a total of 480,000-point clouds with a density of 0.94 pts/m$^3$. The X coordinate range is [-80m, 80m], the Y range is [-80m, 80m], and the Z range is [0m, 20m]. Fig. 5(b) shows the filtering results by the proposed PCAAC filter. In the process of point cloud filtering, low-density sampling points such as street lamps, telegraph poles, power lines and building walls would often be falsely removed by traditional filtering algorithms as noise points, greatly compromising the precision and recall. This problem can be solved by the proposed PCAAC method.

Take a certain region in the original point cloud image as an example. As shown in Fig. 6(a), this image contains 32,839 points with a density of 2.74 pts/m$^3$, including two buildings, a street lamp in the middle, a large grassland, a tree and scattered noise points. Fig. 6(b) shows the filtering result of the statistical outlier removal filter. It can remove outliers but unable to remove noise clusters. Fig. 6(c) shows the filtering result of the radius outlier removal filter. Though it can remove both outliers and noise clusters, it also damaged the environmental features by falsely removing the street lamp in the middle as noises. Fig. 6(d) shows the filtering result of the proposed PCAAC method. With the new method, most noise points were removed with details in environmental features preserved.

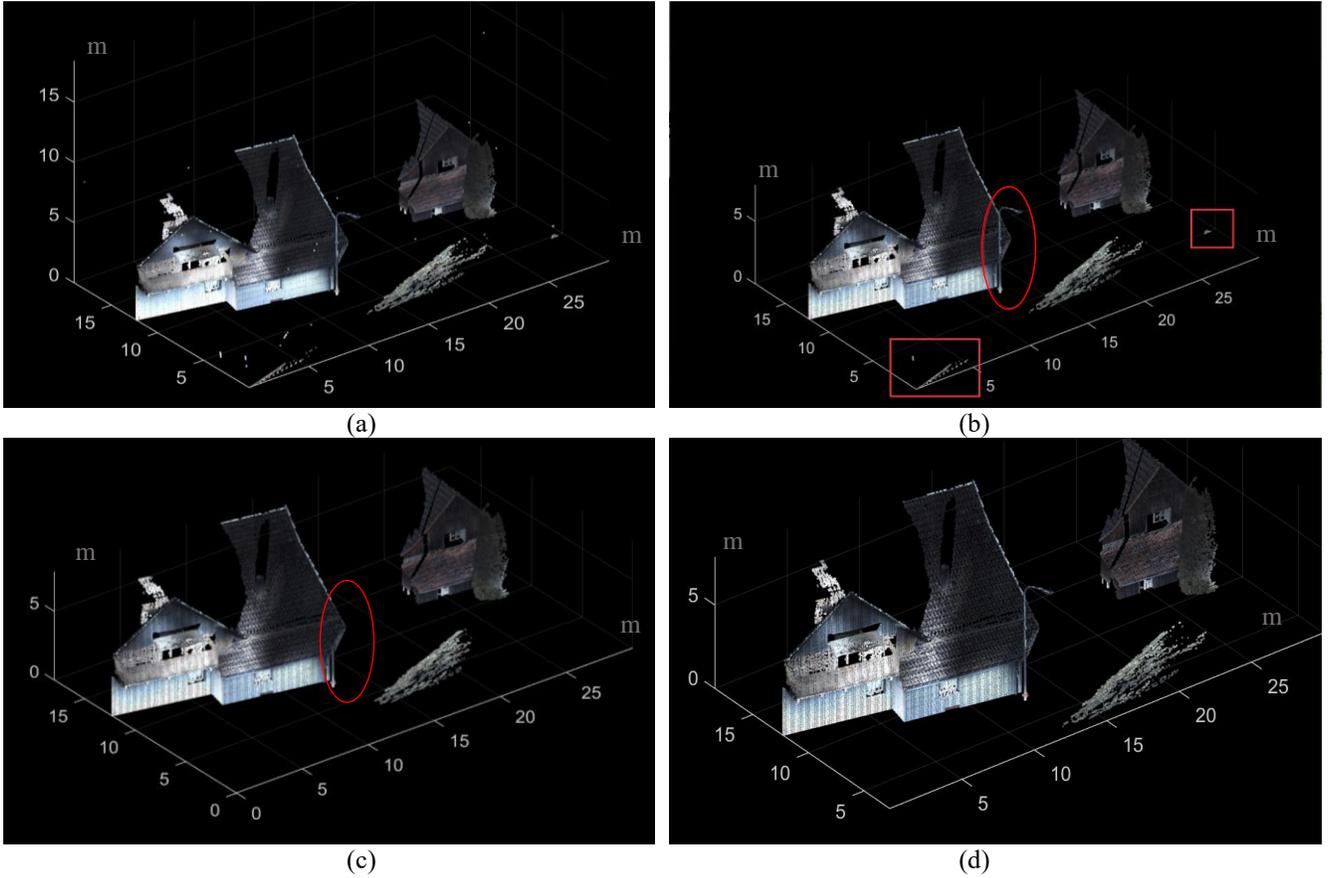

Fig. 6. Original point cloud image and denoised images by different algorithms. (a) Original point cloud image, (b) filtered image by statistical outlier removal filter, (c) filtered image by radius outlier removal filter and (d) filtered image by the proposed PCAAC filter.

In order to evaluate the effectiveness of the filter, four basic parameters and five performance evaluation indicators were used. The four basic parameters are *true positive (TP), false positive (FP), true negative (TN) and false negative (FN)* [1]. The five performance evaluation indicators are *accuracy*, *error*, *precision*, *recall* and $F_1$ value, which are defined as follows:

$$Accuracy = \frac{TP + TN}{TP + FP + TN + FN} \quad (13)$$

$$Error = \frac{FP + FN}{TP + FP + TN + FN} \quad (14)$$

$$Precision = \frac{TP}{TP + FP} \quad (15)$$

$$Recall = \frac{TP}{TP + FN} \quad (16)$$

$$F_1 = \frac{2 \times Precision \times Recall}{Precision + Recall}. \quad (17)$$

The five performance indicators for different filtering methods are compared in Table II.

The statistical outlier removal filter is not effective in removing noise clusters, resulting in a very low recall rate. The two-stage statistical outlier removal filter has a higher recall rate, but its $F_1$ is still outperformed by the radius outlier removal filter. The main problem of the radius outlier removal filter is that useful environmental features (such as street lamps) was removed as noise points, resulting in very low precision and low $F_1$ value. Density-based clustering method yielded significantly higher precision and $F_1$ value than traditional methods. It combines the advantages of statistical outlier removal filter and radius outlier removal filter in removing noise points while retaining useful environmental features. The proposed PCAAC method yielded better outcomes than the existing density-based clustering method. It features high precision and recall rate, and its $F_1$ value reached 0.92. Compared with statistical outlier removal filter, two-stage statistical outlier removal filter, radius outlier removal filter and density-based clustering method, $F_1$ value of the PCAAC method was increased by 135%, 53%, 43% and 15% respectively.

TABLE II
PERFORMANCE EVALUATION OF DIFFERENT METHODS

| Filter | Accuracy | Error | Precision | Recall | $F_1$ |
|---|---|---|---|---|---|
| Statistical Outlier Removal Filter | 99.42% | 0.58% | 58.82% | 28.85% | 0.39 |
| Two-stage Statistical Outlier Removal Filter | 99.52% | 0.48% | 64.29% | 56.25% | 0.60 |
| Radius Outlier Removal Filter | 99.28% | 0.72% | 46.95% | 100% | 0.64 |
| Density-based clustering method | 99.68% | 0.32% | 68.03% | 96.15% | 0.80 |
| PCAAC method | 99.89% | 0.11% | 97.27% | 86.00% | 0.92 |

## V. Conclusion

Point cloud filtering is one of the key steps during lidar signal processing, since noise reduction is essential to obtaining high-quality point cloud data. An adaptive clustering filtering algorithm based on principal component analysis is proposed in this paper. Compared with previous methods, the proposed PCAAC method has high precision, recall and F1 value. Compared with statistical outlier removal filter and radius outlier removal filter, the F-score of the proposed PCAAC filter increased by 135% and 43% respectively with complexity reduced respectively by over 10 times and three times. Compared with the density-based clustering method, the F-score of the proposed PCAAC filter increased by 15% with complexity lowered by 50%. Most noises can be removed while details in environmental features are retained, yielding satisfactory point cloud images. With dimension reduction through PCA, the computational complexity and the running time of the algorithm can be significantly reduced, ensuring better real-time performance.